\documentclass[a4paper,conference]{IEEEtran}
\usepackage[latin9]{inputenc}
\usepackage{amsthm}
\usepackage{amsmath}
\usepackage{graphicx}

\makeatletter

\theoremstyle{plain}

\theoremstyle{remark}
\newtheorem{rem}{\protect\remarkname}
\theoremstyle{plain}
\newtheorem{prop}{\protect\propositionname}
\theoremstyle{plain}
\newtheorem{cor}{\protect\corollaryname}
\theoremstyle{definition}
\newtheorem{example}{\protect\examplename}

\makeatother

\providecommand{\corollaryname}{Corollary}
\providecommand{\examplename}{Example}
\providecommand{\propositionname}{Proposition}
\providecommand{\remarkname}{Remark}
\providecommand{\theoremname}{Theorem}

\begin{document}
\sloppy

%% Paper Title
%% You can use linebreaks \\ within to get better formatting as
%% desired.

\title{Source Coding with in-Block Memory and Causally Controllable Side
Information}
%% Author names and affiliations:
%%
%% Avoiding spaces at the end of the author lines is not a problem with
%% conference papers because we don't use \thanks or \IEEEmembership.
%%
%% For several authors with only one affiliation:
%%
% \author{
%   \IEEEauthorblockN{Hui-Ting Chang and Stefan M.~Moser}
%   \IEEEauthorblockA{Department of Electrical and Computer Engineering\\
%     National Chiao Tung University (NCTU)\\
%     Hsinchu, Taiwan\\
%     Email: \{email-of-hui-ting,email-of-stefan\}@ieee.org}
% }
%%
%% For up to three affiliations:
%%

\author{\IEEEauthorblockN{Osvaldo Simeone \IEEEauthorblockA{New Jersey
Institute of Technology \\
 Email: osvaldo.simeone@njit.edu}}}

\maketitle

\begin{abstract}
The recently proposed set-up of source coding with a side information
``vending machine'' allows the decoder to select actions in order
to control the quality of the side information. The actions can depend
on the message received from the encoder and on the previously measured
samples of the side information, and are cost constrained. Moreover,
the final estimate of the source by the decoder is a function of the
encoder's message and depends causally on the side information sequence.
Previous work by Permuter and Weissman has characterized the rate-distortion-cost
function in the special case in which the source and the ``vending
machine'' are memoryless. In this work, motivated by the related
channel coding model introduced by Kramer, the rate-distortion-cost
function characterization is extended to a model with in-block memory.
Various special cases are studied including block-feedforward and
side information repeat request models.

\end{abstract}

\section{Introduction and System Model}

\begin{figure}[!t]
\centering \includegraphics[bb=136bp 524bp 445bp 642bp,clip,width=3.5in]{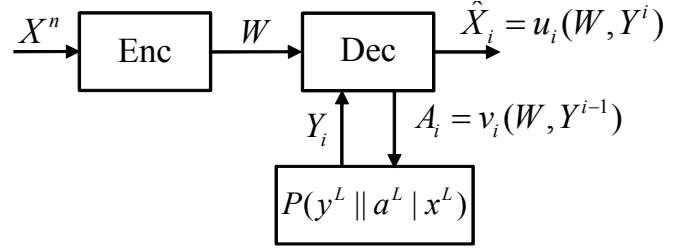}
\caption{Source coding with in-block memory (iBM) and causally controllable
side information.}

\label{figwz}
\end{figure}

Consider the problem of source coding with controllable side information
illustrated in Fig. \ref{figwz}. The encoder compresses a source
$X^{n}=[X_{1},...,X_{n}]$ to a message $W$ of $R$ bits per source
symbol. The decoder, based on the message $W$, takes actions $A_{i}$
for all $i=1,...,n$, so as to control in a causal fashion the measured
side information sequence $Y^{n}$. The action $A_{i}$ is allowed
to be a function of previously measured values $Y^{i-1}$ of the side
information, and the final estimate $\hat{X}_{i}$ is obtained by
the decoder based on message $W$ and as a causal function on the
side information samples. The problem of characterizing the set of
achievable tuples of rate $R$, average distortion $D$ and average
action cost $\Gamma$ was solved in \cite[Sec. II.E]{vending machine}
under the assumptions of a memoryless source $X^{n}$ and of a memoryless
probabilistic model for the side information $Y^{n}$ when conditioned
on the source and the action sequences%
\footnote{The mentioned characterization in \cite[Sec. II.E]{vending machine}
generalizes the result in \cite[Sec. II]{El Gamal} which is restricted
to a model with action-independent side information. %
}. The distribution of the side information sequence given the source
and action sequences is referred to as side information ``vending
machine'' in \cite{vending machine}.

In this work, we generalize the characterization of the rate-distortion-cost
performance for the set-up in Fig. \ref{figwz}, from the memoryless
scenario treated in \cite{vending machine}, to a model in which source
and side information ``vending machine'' have in-block memory (iBM).
With iBM, the probabilistic models for source and ``vending machine''
have memory limited to blocks of size $L$ samples, where $L$ does
not grow with the coding length $n$, as detailed below. The model
under study is motivated by channel coding scenario put forth in \cite{Kramer ITW}
and can be considered to be the source coding counterpart of the latter.

\emph{Notation}: We write $[a,b]=[a,a+1,...,b]$ for integers $b>a$;
$[a,b]=a$ if $a=b$; and $[a,b]$ is empty otherwise. For a sequence
of scalars $x_{1},...,x_{n}$, we write $x^{n}=[x_{1},...,x_{n}]$
and $x^{0}$ for the empty vector. The same notation is used for sequences
of random variables $X^{n}=[X_{1},...,X_{n}]$, or sets $\mathcal{X}^{n}=[\mathcal{X}_{1},...,\mathcal{X}_{n}]$.

\subsection{System Model\label{sec:Model}}

\begin{figure}[!t]
\centering \includegraphics[clip,width=1in]{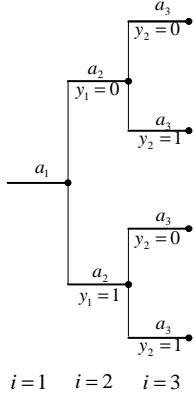} \caption{An \emph{action codetree} $\mathbf{v}^{n}(w,\cdot)$ for a given message
$w\in[1,2^{nR}]$ ($\mathcal{Y}_{i}=\{0,1\}$, $n=3$). }

\label{figcv}
\end{figure}

The system, illustrated in Fig. 1, is described by the following random
variables.
\begin{itemize}
\item A\textbf{ source} $X^{n}$ with iBM of length $L$. The source $X^{n}$
consists of $m$ blocks
\begin{equation}
X_{i}^{L}=(X_{(i-1)L+1},...,X_{(i-1)L+L})
\end{equation}
with $i\in[1,m]$, each of $L$ symbols, so that $n=mL$. The alphabet
is possibly changing across each $L$-block, that is, we have $X_{i}\in\mathcal{X}_{t(i)+1}$,
for $L$ alphabets $\mathcal{X}_{1},...,\mathcal{X}_{L}$, where we
have defined
\begin{equation}
t(i)=r(i-1,L),
\end{equation}
with \emph{$r(x,y)$} being the remainder of $x$ divided by $y$.
\item A \textbf{message} $W\in[1,2^{nR}]$ with $R$ being the rate measured
in bits per source symbol.
\item An\textbf{ action sequence} $A^{n}$ with $A_{i}\in\mathcal{A}_{t(i)+1}$
for $L$ alphabets $\mathcal{A}_{1},...,\mathcal{A}_{L}$.
\item A \textbf{side information sequence} $Y^{n}$ with $Y_{i}\in\mathcal{Y}_{t(i)+1}$
for $L$ alphabets $\mathcal{Y}_{1},...,\mathcal{Y}_{L}$.
\item A\textbf{ source estimate} $\hat{X}^{n}$ with $\hat{X}_{i}\in\mathcal{\hat{X}}_{t(i)+1}$
for $L$ alphabets $\mathcal{\hat{X}}_{1},...,\mathcal{\hat{X}}_{L}$.
\end{itemize}
In order to simplify the notation, in the following, we will write
$\mathcal{X}_{i}$ to denote $\mathcal{X}_{t(i)+1}$ also for $i>L$,
and similarly for the alphabets $\mathcal{A}_{i}$, $\mathcal{Y}_{i}$
and $\mathcal{\hat{X}}_{i}$. The variable are related as follows.
\begin{itemize}
\item The \emph{source} $X^{n}$ has iBM of length $L$ in the sense that
it is characterized as
\begin{equation}
X_{i}=f_{t(i)+1}(Z_{\left\lceil i/L\right\rceil }),\label{eq:X}
\end{equation}
for some functions $f_{i}:\mathcal{Z\rightarrow X}_{i}$, with $i\in[1,L]$,
where $Z_{i}$, with $i\in[1,m]$, is a memoryless process with probability
distribution $P(z)$. Note that (\ref{eq:X}) is equivalent to the
condition that the distribution $P(x^{n})$ factorizes as $\prod_{i=1}^{m}P(x_{i}^{L})$.
\item The \emph{encoder} maps the source $X^{n}$ into a message $W\in[1,2^{nR}]$
according to some function $h:\mathcal{X}^{n}\rightarrow[1,2^{nR}]$
as $W=h(X^{n})$. To denote functional, rather than more general probabilistic,
conditional dependence, we use the notation $1(W|X^{n})$.
\item The \emph{decoder} observes the message $W$ and takes \emph{actions}
$A^{n}$ based also on the observation of the past samples of the
side information sequence. Specifically, for each symbol $i\in[1,n]$
the action $A_{i}$ is selected as
\begin{equation}
A_{i}=v_{i}(W,Y^{i-1}),\label{eq:codetree}
\end{equation}
for some functions $v_{i}:[1,2^{nR}]\times\mathcal{Y}^{i-1}\rightarrow\mathcal{A}_{i}$.
This conditional functional dependence is denoted as $1(a_{i}|\mathbf{v}^{i},y^{i-1})$,
where $\mathbf{v}^{n}\mathbf{=v}^{n}(w,\cdot)$ represents the \emph{action
codetree} (or action strategy) for a given message $w\in[1,2^{nR}]$
in the time interval $i\in[1,n]$, that is, the collection of functions
$v_{i}(w,\cdot)$ in (\ref{eq:codetree}) for all $i\in[1,n]$. A
codetree $\mathbf{v}^{n}(w,\cdot)$ is illustrated in Fig. \ref{figcv}
for $\mathcal{Y}_{i}=\{0,1\}$ and $n=3$. Note that the subtrees
$\mathbf{v}^{i}(w,\cdot)$ with any $i\in[1,n]$ can also be obtained
from Fig. \ref{figcv}.
\end{itemize}

\begin{figure}[!t]
\centering \includegraphics[bb=105bp 107bp 302bp 550bp,clip,width=1in]{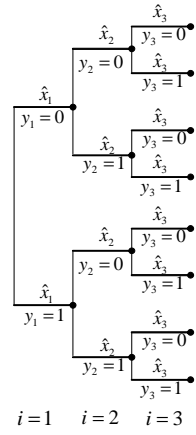}
\caption{A \emph{decoder codetree} $\mathbf{u}^{n}(w,\cdot)$ for a given message
$w\in[1,2^{nR}]$ ($\mathcal{Y}_{i}=\{0,1\}$, $n=3$). }

\label{figcu}
\end{figure}

\begin{itemize}
\item The \emph{side information} has iBM of length $L$ in the sense that
it is generated as a function of the previous actions taken \emph{in
the same block} and of the variable $Z_{\left\lceil i/L\right\rceil }$
(cf. (\ref{eq:X})) as follows
\begin{equation}
Y_{i}=g_{t(i)+1}(A_{i-t(i)},...,A_{i},Z_{\left\lceil i/L\right\rceil }),\label{eq:Ymodel}
\end{equation}
for some functions $g_{i}:\mathcal{A}^{i}\times\mathcal{Z\rightarrow Y}_{i}$,
with $i\in[1,L]$. Note that, as a special case, if the functions
$g_{i}$ do not depend on the actions, equations (\ref{eq:X}) and
(\ref{eq:Ymodel}) imply that the sequences $X^{n}$ and $Y^{n}$
are $L$-block memoryless in the sense that their joint distribution
factorizes as $\prod_{i=1}^{m}P(x_{i}^{L},y_{i}^{L})$.
\item The \emph{decoder}, based on the received message $W$ along with
the current and past samples of the side information sequence, produces
the estimated sequence $\hat{X}^{n}$. Specifically, at each symbol
$i\in[1,n]$, the estimate $\hat{X}_{i}$ is selected as
\begin{equation}
\hat{X}_{i}=u_{i}(W,Y^{i})\label{eq:codetree1}
\end{equation}
for some functions $u_{i}:[1,2^{nR}]\times\mathcal{Y}^{i}\rightarrow\mathcal{\hat{X}}_{i}$.
This conditional functional dependence is denoted as $1(\hat{x}_{i}|\mathbf{u}^{i},y^{i})$,
where $\mathbf{u}^{n}(w,\cdot)$ represents the \emph{decoder codetree}
(or decoder strategy) for a given message $w\in[1,2^{nR}]$ in the
time interval $i\in[1,n]$, that is, the collection of functions $u_{i}(w,\cdot)$
in (\ref{eq:codetree1}) for all $i\in[1,n]$. A codetree $\mathbf{u}^{n}(w,\cdot)$
(along with the subtrees $\mathbf{u}^{i}(w,\cdot)$ with $i\in[1,n]$)
is illustrated in Fig. \ref{figcu} for $\mathcal{Y}_{i}=\{0,1\}$
and $n=3$.
\end{itemize}

\begin{figure}[!t]
\centering \includegraphics[bb=43bp 29bp 464bp 370bp,clip,width=3.5in]{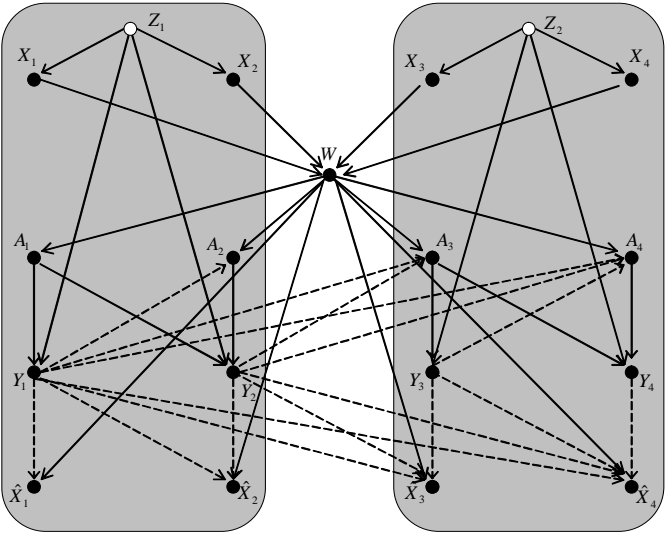}
\caption{FDG for a source coding problem with iBM of length $L=2$ and $n=4$
source symbols (and hence $m=2$ blocks). The two blocks are shaded
and the functional dependence on the side information is drawn with
dashed lines.}

\label{fig0}
\end{figure}

Overall, the probability distribution of the random variables ($X^{n},\mathbf{V}^{n},A^{n},$
$\mathbf{U}^{n},Y^{n},\hat{X}^{n}$) factorizes as
\begin{eqnarray}
\left[\prod_{i=1}^{m}P(x_{i}^{L})\right]P(\mathbf{v}^{n},\mathbf{u}^{n}|x^{n})1(a^{n}||\mathbf{v}^{n},0y^{n-1})\label{eq:jointdistr}\\
\cdot1(\hat{x}^{n}||\mathbf{u}^{n},y^{n})\left[\prod_{i=1}^{m}P(y_{i}^{L}||a_{i}^{L}|x_{i}^{L})\right],\nonumber
\end{eqnarray}
where we have used the \emph{directed conditioning} notation in \cite{Kramer thesis}.
Accordingly, we have defined

\begin{equation}
P(y^{L}||a^{L}|x^{L})=\prod_{i=1}^{L}P(y_{i}|a^{i},x^{L})
\end{equation}
and similarly for the deterministic conditional relationships
\begin{equation}
1(a^{n}||\mathbf{v}^{n},0y^{n-1})=\prod_{i=1}^{n}1(a_{i}|\mathbf{v}^{i},y^{i-1})
\end{equation}
and
\begin{equation}
1(\hat{x}^{n}||\mathbf{u}^{n},y^{n})=\prod_{i=1}^{n}1(\hat{x}_{i}|\mathbf{u}^{i},y^{i}).
\end{equation}
A function dependence graph (FDG) (see, e.g., \cite{Kramer thesis})
illustrating the joint distribution (\ref{eq:jointdistr}) for $L=2$
and $n=2$ (and thus $m=2$) is shown in Fig. \ref{fig0}.

\begin{rem}
In (\ref{eq:jointdistr}), functions $1(a^{n}||\mathbf{v}^{n},0y^{n-1})$
and $1(\hat{x}^{n}||\mathbf{u}^{n},y^{n})$ are fixed as they represent
the map from the branches of the codetrees $\mathbf{v}^{n}$ and $\mathbf{u}^{n}$
as indexed by the side information sequence to the action $a_{i}$
and estimate $\hat{x}_{i}$ as illustrated in Fig. \ref{figcv} and
Fig. \ref{figcu}, respectively.
\end{rem}
Fix a a non-negative and bounded function $d^{L}(x^{L},\hat{x}^{L})$
with domain $\mathcal{X}^{L}\times\mathcal{\hat{X}}^{L}$ to be the
distortion metric and a non-negative and bounded function $\gamma^{L}(a^{L},x^{L})$
with domain $\mathcal{A}^{L}\times\mathcal{\hat{X}}^{L}$ to be the
action cost metric. Under the selected metrics, a triple $(R,D,\Gamma)$
is said to be achievable with distortion $D$ and cost constraint
$\Gamma$, if, for all sufficiently large $m$, there exist codetrees
such that
\begin{equation}
\frac{1}{mL}\sum_{i=1}^{m}E[d^{L}(X_{i}^{L},\hat{X}_{i}^{L})]\leq D+\epsilon\label{eq:conDpf}
\end{equation}
and
\begin{equation}
\frac{1}{mL}\sum_{i=1}^{m}E[\gamma^{L}(A_{i}^{L},X_{i}^{L})]\leq\Gamma+\epsilon\label{eq:conGpf}
\end{equation}
for any $\epsilon>0$. The rate-distortion-cost function $R(D,\Gamma)$
is the infimum of all achievable rates with distortion $D$ and cost
constraint $\Gamma$.
\begin{rem}
\label{rem:specialcase}The system model under study reduces to that
investigated in \cite[Sec. II.E]{vending machine} for the special
case with memoryless sources, i.e., with $L=1$.
\end{rem}

\section{Main Results}

In this section, the rate-distortion-cost function $R(D,\Gamma)$
is derived and some of its properties are discussed. The next section
illustrates various special cases and connections to previous works.

\subsection{Equivalent Formulation}

\begin{figure}[!t]
\centering \includegraphics[bb=47bp 109bp 303bp 550bp,clip,width=1.5in]{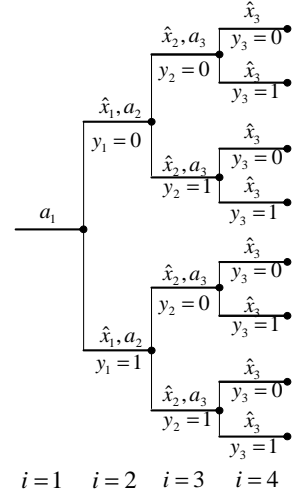}
\caption{A codetree $\mathbf{j}^{n+1}(w,\cdot)$ for a given message $w\in[1,2^{nR}]$
($\mathcal{Y}_{i}=\{0,1\}$, $n=3$).}

\label{figcw}
\end{figure}

We start by showing that the problem can be formulated in terms of
a single codetree. This contrasts with the more natural definitions
given in the previous section, in which two separate codetrees, namely
$\mathbf{v}^{n}(w,\cdot)$ and $\mathbf{u}^{n}(w,\cdot)$, were used
(see Fig. \ref{figcv} and Fig. \ref{figcu}). Towards this end, we
define a ``joint'' codetree $\mathbf{j}^{n+1}(w,\cdot)=(\mathbf{j}^{1}(w,\cdot),...,\mathbf{j}^{n+1}(w,\cdot)$)
that satisfies the functional dependencies
\begin{equation}
1(a_{i}|\mathbf{j}^{i},y^{i-1})=1(a_{i}|\mathbf{v}^{i},y^{i-1}),
\end{equation}
and
\begin{equation}
1(\hat{x}_{i}|\mathbf{j}^{i+1},y^{i})=1(\hat{x}_{i}|\mathbf{u}^{i},y^{i})
\end{equation}
for all $i\in[1,n]$. The codetree $\mathbf{j}^{n+1}(w,\cdot)$ is
illustrated in Fig. \ref{figcw} for $n=3$. Note that the subtree
$\mathbf{j}^{1}(w,\cdot)$ only specifies the action $a_{1}$ to be
taken at time $i=1$, while the the leaves of the tree $\mathbf{j}^{n+1}(w,\cdot)$
are indexed solely by the estimated value $\hat{x}_{n}$.

With this definition, from (\ref{eq:jointdistr}), the probability
distribution of the random variables ($X^{n},\mathbf{J}^{n+1},$ $A^{n},Y^{n},\hat{X}^{n}$)
factorizes as
\begin{eqnarray}
\left[\prod_{i=1}^{m}P(x_{i}^{L})\right]P(\mathbf{j}^{n+1}|x^{n})1(a^{n}||\mathbf{j}^{n},0y^{n-1})\label{eq:jointdistr-1}\\
\cdot1(\hat{x}^{n}||\mathbf{j}_{2}^{n+1},y^{n})\left[\prod_{i=1}^{m}P(y_{i}^{L}||a_{i}^{L}|x_{i}^{L})\right],\nonumber
\end{eqnarray}
where we recall that we have $1(\hat{x}^{n}||\mathbf{j}_{2}^{n+1},y^{n})=\prod_{i=1}^{n}1(\hat{x}_{i}|\mathbf{j}^{i+1},y^{i})$.

\subsection{Rate-Distortion-Cost Function}

\begin{figure}[!t]
\centering \includegraphics[clip,width=3.3in]{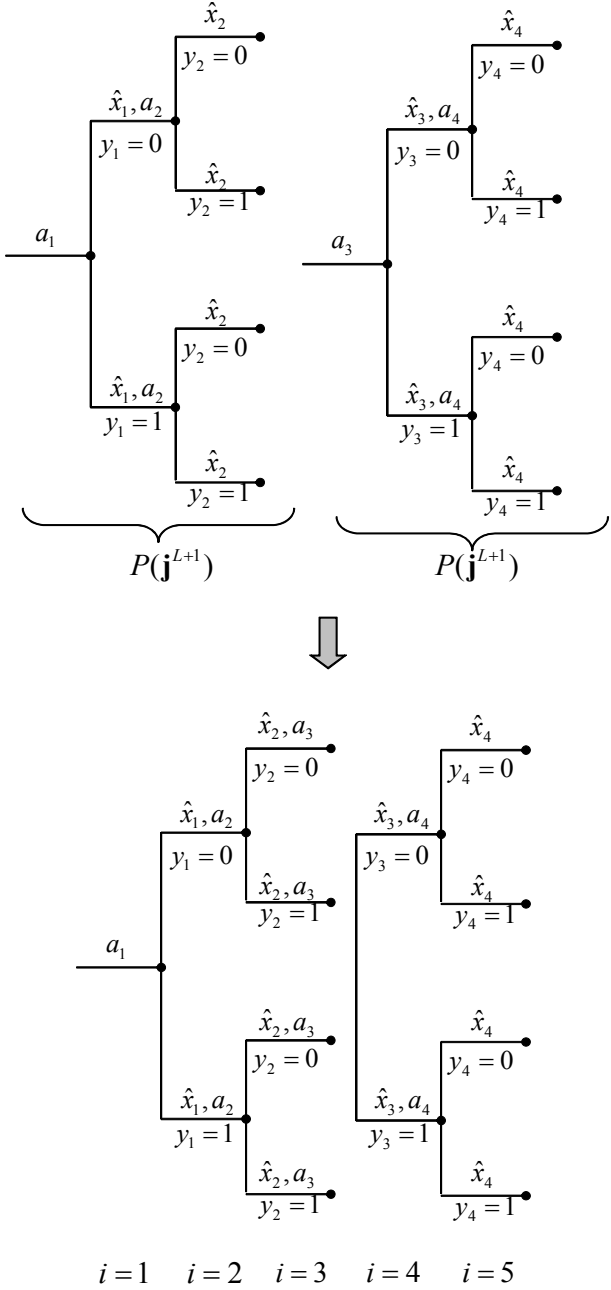} \caption{Illustration of the achievable scheme used in the proof of Proposition
1 for binary alphabets $\mathcal{Y}_{i}=\{0,1\}$ with $m=2$ and
$L=2$. In the top figure, the codetrees $\mathbf{j}_{i}^{3}(w,\cdot)$
for $i=1,2$, which are generated i.i.d. with probability $P(\mathbf{j}^{L+1}$),
are depicted. In the bottom figure, the resulting codetree $\mathbf{j}^{5}(w,\cdot)$
is shown. It is noted that the action $a_{3}$ in the codetree $\mathbf{j}^{5}(w,\cdot)$
in the bottom figure is obtained from the codetree $\mathbf{j}_{i}^{3}(w,\cdot)$
with $i=2$ on the top, and is thus independent of the value of $y^{2}$.}

\label{figcw-1}
\end{figure}

Using the representation in terms of a single codetree given above,
we now provide a characterization of the rate-distortion-cost function.
\begin{prop}
\label{prop:The-rate-distortion-cost-functio}The rate-distortion-cost
function is given by
\begin{equation}
R(D,\Gamma)=\frac{1}{L}\min I(X^{L};\mathbf{J}^{L+1})\label{eq:rdc1}
\end{equation}
where the joint distribution of the variables \textup{$X^{L}$,$Y^{L}$,$A^{L}$,$\hat{X}^{L}$
and of the codetree $\mathbf{J}^{L+1}$ factorizes as}
\begin{eqnarray}
P(x^{L})P(\mathbf{j}^{L+1}|x^{L})1(a^{L}||\mathbf{j}^{L},0y^{L-1})\label{eq:distrprop}\\
\cdot1(\hat{x}^{L}||\mathbf{j}_{2}^{L+1},y^{L})P(y^{L}||a^{L}|x^{L}),\nonumber
\end{eqnarray}
and the minimization is performed over the conditional distribution
$P(\mathbf{j}^{L+1}|x^{L})$ of the codetree under the constraints
\begin{equation}
\frac{1}{L}E[d^{L}(X^{L},\hat{X}^{L})]\leq D\label{eq:constD}
\end{equation}
and
\begin{equation}
\frac{1}{L}E[\gamma^{L}(A^{L},X^{L})]\leq\Gamma.\label{eq:constGamma}
\end{equation}
\end{prop}
\begin{IEEEproof}
The achievability of Proposition \ref{prop:The-rate-distortion-cost-functio}
follows from classical random coding arguments. Specifically, the
encoder draws the codetrees $\mathbf{j}^{n+1}(w,\cdot)$ for all $w\in[1,2^{n(R(D)+\delta)}${]}
with some $\delta>0$, as follows. First, for each $w\in[1,2^{n(R(D)+\delta)}${]}
a concatenation of $m$ codetrees $\mathbf{j}_{i}^{L+1}(w,\cdot)$
of length $L+1$, with $i\in[1,m]$, is generated, such that the constituent
codetrees $\mathbf{j}_{i}^{L+1}(w,\cdot)$ are i.i.d. and distributed
with probability $P(\mathbf{j}^{L+1}$). The codetree $\mathbf{j}^{n+1}(w,\cdot)$
is then obtained by combining the leaves and the root of successive
constituent codetrees: the leaves of the past codetree specify the
estimates for the previous time instant, while the root of the next
codetree specify the action for the current time instant. The procedure
is illustrated in Fig. \ref{figcw-1}.

Encoding is performed by looking for a message $w\in[1,2^{n(R(D)+\delta)}${]}
such that the corresponding pair ($x^{n},\mathbf{j}^{n+1}(w,$$\cdot$))
is (strongly) jointly typical with respect to the joint distribution
$P(x^{L})P(\mathbf{j}^{L+1}|x^{L})$, when the sequences ($x^{n},\mathbf{j}^{n+1}(w,$$\cdot$))
are seen as the memoryless $m$-sequences ($x_{1}^{L},\mathbf{j}_{1}^{L+1}(w,$$\cdot$)),...,($x_{m}^{L},\mathbf{j}_{m}^{L+1}(w,$$\cdot$)).
By the covering lemma \cite[Lemma 3.3]{El Gamal Kim}, rate $1/L\cdot I(X^{L};\mathbf{J}^{L+1})$
suffices to guarantee the reliability of this step. Moreover, if the
distribution $P(\mathbf{j}^{L+1}|x^{L})$ is selected so as to satisfy
(\ref{eq:constD}) and (\ref{eq:constGamma}), then, by the typical
average lemma \cite{El Gamal Kim}, the constraints (\ref{eq:conDpf})
and (\ref{eq:conGpf}) are also guaranteed to be met for sufficiently
large $n$. The proof of the converse can be found in Appendix \ref{sec:Proof-of-the}.\end{IEEEproof}
\begin{rem}
\label{rem:The-rate-distortion-cost-functio}The rate-distortion-cost
function can also be expressed in terms of two separate codetrees
using the definitions given in Sec. \ref{sec:Model}. Specifically,
following similar steps as in the proof of Proposition \ref{prop:The-rate-distortion-cost-functio},
the rate-distortion-cost function can be expressed as the minimization
\begin{equation}
R(D,\Gamma)=\frac{1}{L}\min I(X^{L};\mathbf{V}^{L},\mathbf{U}^{L})\label{eq:rdcsingletree}
\end{equation}
where the joint distribution of the variables $X^{L}$,$Y^{L}$,$A^{L}$,$\hat{X}^{L}$
and of the codetrees $\mathbf{V}^{L}$ and $\mathbf{U}^{L}$ factorizes
as
\begin{eqnarray}
P(x^{L})P(\mathbf{v}^{L},\mathbf{u}^{L}|x^{L})1(a^{L}||\mathbf{v}^{L},0y^{L-1})\label{eq:-1}\\
\cdot1(\hat{x}^{L}||\mathbf{u}^{L},y^{L})P(y^{L}||a^{L}|x^{L}),\nonumber
\end{eqnarray}
and the minimization is performed over the conditional distribution
$P(\mathbf{v}^{L},\mathbf{u}^{L}|x^{L})$ of the codetrees under the
constraints (\ref{eq:constD}) and (\ref{eq:constGamma}).
\end{rem}

\begin{rem}
The rate-distortion-cost function in Proposition \ref{prop:The-rate-distortion-cost-functio}
does not include auxiliary random variables, since the codetree $\mathbf{J}^{L+1}$
is part of the problem specification. This is unlike the characterization
given in \cite{vending machine} for the memoryless case. Moreover,
problem (\ref{eq:rdc1}) is convex in the unknown $P(\mathbf{j}^{L+1}|x^{L})$
and hence can be solved using standard algorithms. It is also noted
that, extending \cite{Dupuis}, one may devise a Blahut-Arimoto-type
algorithm for the calculation of the rate-distortion-cost function.
This aspect is not further investigated here.
\end{rem}
Based on the definition of $\mathbf{J}^{L+1}$, we have the following
cardinality bound on the number of codetrees to be considered in the
optimization (\ref{eq:rdc1}):
\begin{equation}
|\mathcal{J}^{L+1}|\leq|\mathcal{A}_{1}||\mathcal{\hat{X}}_{L}|^{|\mathcal{Y}^{L}|}\prod_{i=1}^{L-1}(|\mathcal{\hat{X}}_{i}||\mathcal{A}_{i+1}|)^{|\mathcal{Y}^{i}|}.\label{eq:cardbound1}
\end{equation}
The following lemma shows that the this cardinality bound can be improved.
\begin{cor}
\label{cor:The-rate-distortion-cost-functio}In the optimization (\ref{eq:rdc1}),
the number of codetrees \textup{$\mathbf{J}^{L+1}$} can be limited
as
\begin{equation}
|\mathcal{J}^{L+1}|\leq|\mathcal{X}^{L}|+3\label{eq:cardbound2}
\end{equation}
without loss of optimality. \end{cor}
\begin{IEEEproof}
See Appendix B.\end{IEEEproof}
\begin{rem}
The achievable scheme used to prove Proposition \ref{prop:The-rate-distortion-cost-functio}
adapts the actions only to the side information samples corresponding
to the same $L$-block. More precisely, the action $A_{i}$ depends,
through the selected codetree, only on the side information samples
$Y_{i-t(i)},...,Y_{i-1}$. Since the problem definition allows, via
(\ref{eq:codetree}), for actions that depend on all past side information
samples, namely $Y^{i-1},$ this result demonstrates that adapting
the actions across the blocks cannot improve the rate-distortion-cost
function. This is consistent with the finding in \cite{vending machine},
where it is shown that adaptive actions do not improve the rate-distortion
performance for a memoryless model, i.e., with $L=1$. Similarly,
one can conclude from Proposition \ref{prop:The-rate-distortion-cost-functio}
that, while adapting the estimate $\hat{X}_{i}$ to the side information
samples within the same $L$-block, namely $Y_{i-t(i)},...,Y_{i}$,
is generally advantageous, adaptation across the blocks is not. This
extends the results in \cite{El Gamal}, in which it is shown that,
for $L=1$, the estimate can depend only on the current value of the
side information without loss of optimality.
\end{rem}

\section{Special Cases and Examples}

In this section, we detail some further consequences of Proposition
\ref{prop:The-rate-distortion-cost-functio} and connections with
previous work.

\subsection{Memoryless Source ($L=1$)}

As mentioned in Remark \ref{rem:specialcase}, if $L=1$, the model
at hand reduces to the standard one with memoryless sources, in which
the joint distribution of $X^{n}$ and $Y^{n}$ factorizes as $\prod_{i=1}^{n}P(x_{i},y_{i})$.
This model was studied in \cite{vending machine}, where the rate-distortion-cost
function was derived. The result in \cite[Sec. II-E]{vending machine}
can be seen to be a special case of Proposition \ref{prop:The-rate-distortion-cost-functio}.

\subsection{Action-Independent Side Information}

Here we consider the case in which the side information is action
independent, that is, we have $P(y^{L}||a^{L}|x^{L})=P(y^{L}|x^{L})$.
Under this assumption, the action sequence does not need to be included
in the model, and, from (\ref{eq:rdcsingletree}), the rate-distortion
function is given by
\begin{equation}
R(D)=\frac{1}{L}\min I(X^{L};\mathbf{U}^{L}),\label{eq:rate-acind-1}
\end{equation}
where the joint distribution of the variables $X^{L}$,$Y^{L}$,$\hat{X}^{L}$
and of the codetree $\mathbf{U}^{L}$ factorizes as
\begin{equation}
P(x^{L})P(\mathbf{u}^{L}|x^{L})1(\hat{x}^{L}||\mathbf{u}^{L},y^{L})P(y^{L}|x^{L}),
\end{equation}
and the minimization is performed over the conditional distribution
$P(\mathbf{u}^{L}|x^{L})$ of the codetrees under the constraint (\ref{eq:constD}).
Note that, given the absence of actions, we have used the formulation
in terms of individual codetrees discussed in Remark \ref{rem:The-rate-distortion-cost-functio}
in order to simplify the notation. Using arguments similar to Corollary
\ref{cor:The-rate-distortion-cost-functio}, one can show that the
size of the codetree alphabet can be limited to $|\mathcal{U}^{L}|\leq|\mathcal{X}^{L}|+2$
without loss of optimality. For $L=1$, the characterization (\ref{eq:rate-acind-1})
reduces to the one derived in \cite[Sec. II]{El Gamal}.

\subsection{Block-Feedforward Model}

As a specific instance of the setting with action-independent side
information, we consider here the \emph{block-feedforward} model in
which we have $Y_{i}=X_{i-1}$ for all $i$ not multiple of $L$ and
$Y_{i}$ equal to a fixed symbol in $\mathcal{Y}_{i}$ otherwise.
This model is related to the feedforward set-up studied in \cite{weissman03,pradhan,Naiss}
with the difference that here feedforward is limited to within the
$L$-blocks. In other words, the side information is $Y_{i}=X_{i-1}$
only if $X_{i-1}$ is in the same $L$-block as $Y_{i}$ and is not
informative otherwise. We now show that, similar to \cite{pradhan},
the rate-distortion function with block-feedforward can be expressed
in terms of directed information and does not entail an optimization
over the codetrees.
\begin{cor}
\textup{\label{cor:For-the-block-feedforward}For the block-feedforward
model, the rate-distortion function is given by }
\begin{equation}
R(D)=\frac{1}{L}\min I(\hat{X}^{L}\rightarrow X^{L})\label{eq:rate-ff}
\end{equation}
where the joint distribution of the variables \textup{$X^{L}$, $Y^{L}$
and $\hat{X}^{L}$ factorizes as}
\begin{equation}
P(x^{L})P(\hat{x}^{L}|x^{L})P(y^{L}|x^{L}),
\end{equation}
and the minimization is performed over the conditional distribution
$P(\hat{x}^{L}|x^{L})$ under the constraint (\ref{eq:constD}). \end{cor}
\begin{rem}
In the feedforward model studied in \cite{weissman03,pradhan,Naiss},
feedforward of the source $X^{n}$ is not restricted to take place
only within the $L$-blocks, namely we have $Y_{i}=X^{i-1}$ for all
$i\in[1,n]$. As a result, the rate-distortion function is proved
in \cite{pradhan,Naiss} to be given by the limit of (\ref{eq:rate-ff})
over $L$.\end{rem}
\begin{IEEEproof}
The achievability is obtained by using concatenated codetrees of length
$L$ similar to Proposition \ref{prop:The-rate-distortion-cost-functio}.
However, unlike Proposition \ref{prop:The-rate-distortion-cost-functio},
the codetrees are generated according to the distribution $p(\hat{x}^{L}||0x^{L-1})$
as done in \cite{pradhan,Naiss}. The proof of achievability is completed
as in \cite{pradhan,Naiss}. As for the converse, starting from (\ref{eq:rate-acind-1}),
we write

\begin{eqnarray}
I(X^{L};\mathbf{U}^{L}) & = & \sum_{i=1}^{L}I(X_{i};\mathbf{U}^{L}|X^{i-1})\nonumber \\
 & = & \sum_{i=1}^{L}I(X_{i};\mathbf{U}^{L},\hat{X}^{i}|X^{i-1})\nonumber \\
 & \geq & \sum_{i=1}^{L}I(X_{i};\hat{X}^{i}|X^{i-1})\nonumber \\
 & = & I(\hat{X}^{L}\rightarrow X^{L}),
\end{eqnarray}
where the second equality follows since $\hat{X}^{i}$ is a function
of the codetree $\mathbf{U}^{L}$ and of $Y^{i}=X^{i-1}$; the inequality
follows by the non-negativity of the mutual information; and the last
equality is a consequence of the definition of directed information
\cite{Kramer thesis}. \end{IEEEproof}
\begin{example}
Consider a binary source with iBM of length $L=2$ and block-feedforward
such that variables $X_{i}$, for all odd $i$, are i.i.d. $Bern(p)$,
with $0\leq p\leq0.5$, while for all even $i$ we have $X_{i}=X_{i-1}\oplus Q_{i}$
with $Q_{i}$ being i.i.d. $Bern(q)$, with $0\leq q\leq0.5$ and
independent of $X_{i}$ for all odd $i$. Assuming Hamming distortion
$d^{2}(x^{2},\hat{x}^{2})=\sum_{i=1}^{2}1(x_{i},\hat{x}_{i})$, from
Corollary \ref{cor:For-the-block-feedforward}, we easily obtain that,
if $D<(p+q)/2$, the rate-distortion function is given as
\begin{equation}
\min_{D_{1}+D_{2}\leq2D}\frac{1}{2}\left[H_{2}(p)-H_{2}(D_{1})+H_{2}(q)-H_{2}(D_{2})\right]
\end{equation}
where the minimization is under the constraints $D_{1}\leq p$ and
$D_{2}\leq q$, and is zero otherwise.
\end{example}

\subsection{Side Information Repeat Request}

Consider the situation in which the decoder at any time $i$, upon
the observation of the side information $Y_{i}$, can decide whether
to take a second measurement of the side information, thus paying
the associated cost, or not. To elaborate, assume a memoryless source
$X^{n}$ with distribution $P(x)$. At any time \emph{$i$}, the first
observation $Y_{i1}$ of the side information is distributed according
to the memoryless channel $P(y_{1}|x)$ when the input is $X_{i}=x$,
while the second observation $Y_{i2}$ depends on the action $A_{i}=a$
via the memoryless channel $P(y_{2}|x,a)$ with input $X_{i}=x$.

\begin{figure}[!t]
\centering \includegraphics[bb=41bp 27bp 468bp 370bp,clip,width=3.5in]{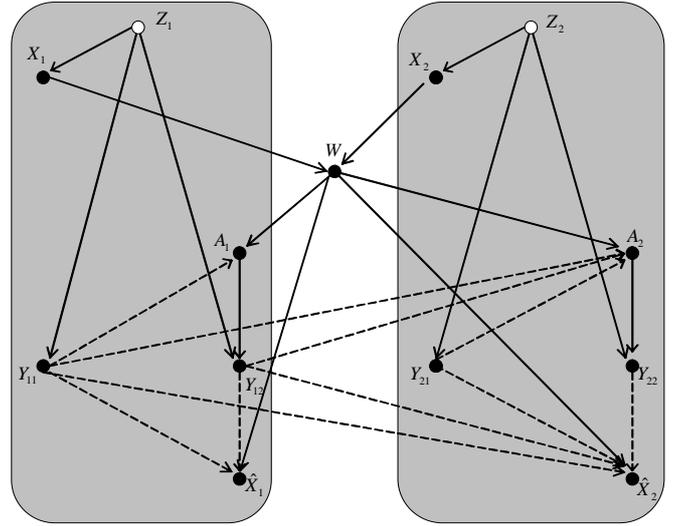}
\caption{FDG for the model with side information repeat request. }

\label{figsl}
\end{figure}

This scenario can be easily seen to be a special case of the model
under study with iBM of size $L=2$. The corresponding FDG is illustrated
in Fig. \ref{figsl}. By comparing this FDG with the general FDG in
Fig. \ref{fig0}, it is seen that the model under study in this section
can be obtained from the one presented in Sec. \ref{sec:Model} by
appropriately setting the alphabets of given subset of variables to
empty sets and by relabeling.

A characterization of the rate-distortion-cost function can be easily
derived as a special case of Proposition \ref{prop:The-rate-distortion-cost-functio}.
Here we focus on a specific simple example. In particular, we assume
that the channel $P(y_{1}|x)$ for the first measurement is an erasure
channel with erasure probability $\epsilon$. Moreover, the channel
$P(y_{2}|x,a)$ for the second measurement is an independent and identical
erasure channel if $a=1$, while it produces $Y_{2}$ equal to the
erasure symbol with probability 1 if $a=0$. In other words, the action
$a=1$ corresponds to performing a second measurement of the side
information over an independent realization of the same erasure channel.

It is apparent that, if $Y_{1}=X$, one can set $A=0$ without loss
of optimality. Instead, if $Y_{1}$ equals the erasure symbol, then,
in the absence of action cost constraints, it is clearly optimal to
set $A=1$. In so doing, the side information channel is converted
into an equivalent erasure channel with erasure probability $\epsilon^{2}$.
Therefore, the rate-distortion is given by \cite{diggavi,weissman verdu}
\begin{equation}
R(D,\Gamma)=\epsilon^{2}\left(1-H_{2}\left(\frac{D}{\epsilon^{2}}\right)\right)\label{eq:rdrr}
\end{equation}
for $D\leq\epsilon^{2}/2$ and zero otherwise, as long as the action
cost budget $\Gamma$ is large enough. More specifically, given the
discussion above, it can be seen that $\Gamma\geq\epsilon$ suffices
to achieve (\ref{eq:rdrr}).

\section{Concluding Remarks}

Models with in-block memory (iBM), first proposed in the context of
channel coding problems in \cite{Kramer ITW} and here for source
coding, provide tractable extensions of standard memoryless models.
Specifically, in this paper, we have presented results for a point-to-point
system with controllable side information at the receiver and iBM.
Interesting generalizations include the investigation of multi-terminal
models.

\section*{Acknowledgment}

{\small The author would like to thank Gerhard Kramer for the very useful
comments and suggestions.}{\small \par}

\appendices{}

\section{Proof of the Converse of Proposition \ref{prop:The-rate-distortion-cost-functio}\label{sec:Proof-of-the}}

For any code achieving rate $R$ with distortion $D$ and cost $\Gamma$,
we have the following series of inequalities:
\begin{eqnarray*}
nR & \geq & H(W)=I(W;X^{n})\\
 & \overset{(a)}{=} & \sum_{i=1}^{m}H(X_{i}^{L})-\sum_{i=1}^{m}H(X_{i}^{L}|X^{(i-1)L},W)\\
 & \overset{(b)}{=} & \sum_{i=1}^{m}H(X_{i}^{L})-\sum_{i=1}^{m}H(X_{i}^{L}|X^{(i-1)L},W,Y^{(i-1)L})\\
 & \overset{(c)}{=} & \sum_{i=1}^{m}H(X_{i}^{L})-\sum_{i=1}^{m}H(X_{i}^{L}|X^{(i-1)L},W,Y^{(i-1)L},\mathbf{\bar{J}}_{i}^{L+1})\\
 & \overset{(d)}{\geq} & \sum_{i=1}^{m}H(X_{i}^{L})-\sum_{i=1}^{m}H(X_{i}^{L}|\mathbf{\bar{J}}_{i}^{L+1})\\
 & \overset{(e)}{=} & mH(X^{L})-mH(X^{L}|\mathbf{\bar{J}}_{i}^{L+1},T)\\
 & \geq & mI(X^{L};\mathbf{\bar{J}}^{L+1}),
\end{eqnarray*}
where (a) follows due to the block memory of the source $X^{n}$;
(b) follows due to the Markov chain $X_{i}^{L}-(X^{(i-1)L},W)-Y^{(i-1)L}$;
(c) is obtained by defining $\mathbf{\bar{J}}_{i}^{L+1}$ as the subtree
of $\mathbf{J}^{i+1}$ corresponding to $Y^{(i-1)L}$, respectively,
and noting that $\mathbf{\bar{J}}_{i}^{L+1}$ is a function of ($W,Y^{(i-1)L}$);
(d) is due to the fact that conditioning cannot increase entropy;
(e) is obtained by defining a random variable $T$ uniformly distributed
in the set $[1,m]$ and independent of all other variables, and also
the variables $\mathbf{\bar{J}}^{L+1}=\mathbf{\bar{J}}_{T}^{L+1}$
and $X^{L}=X_{T}^{L}$, and using the fact that the distribution of
$X_{i}^{L}$ does not depend on $i$.

Given the definitions above, and setting $A^{L}=A_{T}^{L}$, the joint
distribution of the random variables at hand factorizes as
\begin{eqnarray}
P(x^{L})P(\mathbf{\bar{j}}^{L+1}|x^{L})1(a^{L}||\mathbf{\bar{j}}^{L},0y^{L-1})\label{eq:27}\\
\cdot1(\hat{x}^{L}||\mathbf{\bar{j}}_{2}^{L+1},y^{L})P(y^{L}||a^{L}|x^{L}),\nonumber
\end{eqnarray}
where we have defined $P(\mathbf{\bar{j}}^{L+1}|x^{L})=\frac{1}{m}\sum_{t=1}^{m}P(\mathbf{\bar{j}}^{L+1}|x^{L},t)$.
Note that, in showing (\ref{eq:27}), it is critical that, as per
(\ref{eq:Ymodel}), the side information $Y_{i}^{L}$ in the $i$th
block depends only on the actions in the $i$th block. The proof is
concluded by noting that the defined random variables also satisfy
the constraints (\ref{eq:constD}) and (\ref{eq:constGamma}) due
to the fact that any code at hand must satisfy the conditions (\ref{eq:conDpf})
and (\ref{eq:conGpf}), respectively.

\section{Proof of Corollary \ref{cor:The-rate-distortion-cost-functio}}

Assume that a rate is achievable for some distribution $P(\mathbf{j}^{L+1}|x^{L})$,
where the cardinality of $\mathbf{J}^{L+1}$ is limited only by the
count of available codetrees as in (\ref{eq:cardbound1}). We want
to show that the same rate can be achieved by limiting the alphabet
of available codetrees as in (\ref{eq:cardbound2}). To this end,
we first write the joint distribution (\ref{eq:distrprop}) as
\begin{eqnarray}
P(\mathbf{j}^{L+1})P(x^{L}|\mathbf{j}^{L+1})1(a^{L}||\mathbf{j}^{L},0y^{L-1})\label{eq:appdistr}\\
\cdot1(\hat{x}^{L}||\mathbf{j}_{2}^{L+1},y^{L})P(y^{L}||a^{L}|x^{L}).\nonumber
\end{eqnarray}
Now, fix the so obtained distribution $P(x^{L}|\mathbf{j}^{L+1})$
and recall that the other terms in (\ref{eq:appdistr}) are also fixed
by the problem definition. Now, the quantities appearing in Proposition
\ref{prop:The-rate-distortion-cost-functio} can be written as convex
combinations of functions of the terms fixed above, in which the distribution
$P(\mathbf{j}^{L+1})$ defines the coefficients of the combinations.
Specifically, we have: (\emph{i}) the distribution $P(x^{L}$$)=\sum_{\mathbf{j}^{L+1}}P(\mathbf{j}^{L+1})P(x^{L}|\mathbf{j}^{L+1})$
for all $x^{L}\in\mathcal{X}^{L}$ (but one), which fixes $H(X^{L})$;
(\emph{ii}) the conditional entropy $H(X^{L}|\mathbf{J}^{L+1})=\sum_{\mathbf{j}^{L+1}}P(\mathbf{j}^{L+1})H(X^{L}|\mathbf{J}^{L+1}=\mathbf{j}^{L+1})$;
and (\emph{iii}) the averages $E[d^{L}(X^{L},\hat{X}^{L})]$ and $E[\gamma^{L}(A^{L},X^{L})]$.
It follows by the Caratheodory theorem that we can limit the alphabet
of $\mathbf{J}^{L+1}$ as in (\ref{eq:cardbound2}) without loss of
optimality.

{\small %% References:
%% We recommend the usage of BibTeX:
%%
%\bibliographystyle{IEEEtran}
%\bibliography{definitions,bibliofile}
%%
%% where we here have assume the existence of the files
%% definitions.bib and bibliofile.bib.
%% BibTeX documentation can be obtained at:
%% http://www.ctan.org/tex-archive/biblio/bibtex/contrib/doc/
%%
%%
%%
%% Or manual references (pay attention to consistency!):
}
\end{document}